\documentclass[aps,prx,preprint,superscriptaddress,citeautoscript]{revtex4-1}

\usepackage[T1]{fontenc}       
\usepackage{xcolor}
\usepackage{graphicx}
\usepackage{epstopdf}
\usepackage{amssymb,amsmath}
\usepackage{pdfpages}
\usepackage{etoolbox} 
\makeatletter
\patchcmd{\@outputpage@head}{\@ifx{\LS@rot\@undefined}{}{\LS@rot}}{}{}{}
\makeatother

\begin{document}


\title{Stable biexcitons in two-dimensional metal-halide perovskites with strong dynamic lattice disorder}


\author{F\'elix~Thouin}
\affiliation{School of Physics, Georgia Institute of Technology, 837 State Street NW, Atlanta, Georgia 30332, USA}

\author{Stefanie~Neutzner}
\thanks{FT and SN are to be considered first co-authors of this manuscript.}
\affiliation{Center for Nano Science and Technology @Polimi, Istituto Italiano di Tecnologia, via Giovanni Pascoli 70/3, 20133 Milano, Italy}

\author{Daniele~Cortecchia}
\affiliation{Center for Nano Science and Technology @Polimi, Istituto Italiano di Tecnologia, via Giovanni Pascoli 70/3, 20133 Milano, Italy}
\affiliation{ Interdisciplinary Graduate School, Energy Research Institute @ NTU (ERI@N), Nanyang Technological University, Singapore 639798}

\author{Vlad~Alexandru~Dragomir}
\affiliation{D\'epartement de physique, Universit\'e de Montr\'eal, Case Postale 6128, Succursale centre-ville, Montr\'eal, Qu\'ebec H3C~3J7, Canada}

\author{Cesare~Soci}
\affiliation{Centre for Disruptive Photonic Technologies, TPI, SPMS, 21 Nanyang Link, Singapore 637371}

\author{Teddy~Salim}
\affiliation{School of Materials Science and Engineering, Nanyang Technological University, 50 Nanyang Avenue, 639798, Singapore}

\author{Yeng~Ming~Lam}
\affiliation{School of Materials Science and Engineering, Nanyang Technological University, 50 Nanyang Avenue, 639798, Singapore}

\author{Richard~Leonelli}
\affiliation{D\'epartement de physique, Universit\'e de Montr\'eal, Case Postale 6128, Succursale centre-ville, Montr\'eal, Qu\'ebec H3C~3J7, Canada}

\author{Annamaria~Petrozza}
\affiliation{Center for Nano Science and Technology @Polimi, Istituto Italiano di Tecnologia, via Giovanni Pascoli 70/3, 20133 Milano, Italy}

\author{Ajay~Ram~Srimath~Kandada}
\email[]{srinivasa.srimath@iit.it}
\affiliation{School of Physics, Georgia Institute of Technology, 837 State Street NW, Atlanta, Georgia 30332, USA}
\affiliation{Center for Nano Science and Technology @Polimi, Istituto Italiano di Tecnologia, via Giovanni Pascoli 70/3, 20133 Milano, Italy}
\affiliation{School of Chemistry and Biochemistry, Georgia Institute of Technology, 901 Atlantic Drive NW, Atlanta, Georgia 30332, USA}

\author{Carlos~Silva}
\email[]{carlos.silva@gatech.edu}
\affiliation{School of Physics, Georgia Institute of Technology, 837 State Street NW, Atlanta, Georgia 30332, USA}
\affiliation{School of Chemistry and Biochemistry, Georgia Institute of Technology, 901 Atlantic Drive NW, Atlanta, Georgia 30332, USA}


\date{\today}

\begin{abstract}
With strongly bound and stable excitons at room temperature, single-layer, two-dimensional organic-inorganic hybrid perovskites are viable semiconductors for light-emitting quantum optoelectronics applications. In such a technological context, it is imperative to comprehensively explore all the factors --- chemical, electronic and structural --- that govern  strong multi-exciton correlations. Here, by means of two-dimensional coherent spectroscopy, we examine excitonic many-body effects in pure, single-layer (PEA)$_2$PbI$_4$ (PEA = phenylethylammonium). We determine the binding energy of biexcitons --- correlated two-electron, two-hole quasiparticles --- to be $44 \pm 5$\,meV at room temperature. The extraordinarily high values are similar to those reported in other strongly excitonic two-dimensional materials such as transition-metal dichalchogenides. Importantly, we show that this binding energy increases by  $\sim25$\% upon cooling to 5\,K. Our work highlights the importance of multi-exciton correlations in this class of technologically promising, solution-processable materials, in spite of the strong effects of lattice fluctuations and dynamic disorder. 
\end{abstract}


\maketitle

\section{Introduction}
Excitonic interactions in two-dimensional (2D) semiconductors garner considerable attention, both due to their relevance in quantum optoelectronics and to the richness of their physics~\cite{Butler2013, Wang2012}. Whilst their peculiar electronic characteristics have been observed in many single-layer, atomically thin systems, most recently in transition-metal dichalcogenides~\cite{Ugeda2014, He2014, Chernikov2014, Liu2014}, 
two-dimensional hybrid organic-inorganic perovskites (HOIP) offer a valuable alternative test system due to the ability to dramatically alter the structure of the inorganic sub-lattice via choice of the organic component~\cite{Cortecchia2017a}. Specifically, we consider HOIPs that consist of lead-iodide octahedra forming 2D lattice planes, separated by long organic cationic ligands (see inset of Fig.~\ref{absorption_spectrum}), resulting in quantum-well-like structures with strong electronic confinement within the isolated octahedral layers~\cite{Saparov2016}. HOIPs can be readily processed from solution and yet keep the strong excitonic character, even when embodied within polycrystalline films. In that regard, they not only offer an experimentally accessible material system to study 2D many-body exciton physics, but also make a compelling case for novel quantum optoelectronic technologies. In this manuscript, we quantify many-body electron-hole correlations, and we find that these are strong enough to make multi-excitons stable and relevant quasi-particles at ambient conditions, even in the presence of robust energetic disorder.  

Though HOIPs fall in the family of excitonic 2D semiconductors, they exhibit many distinct photophysical characteristics, mainly due to their unique chemical and stuctural composition~\cite{Pedesseau2016}. For example, dielectric confinement arising from the intercalating organic layers increases the Coulomb correlations substantially, resulting in a strong increase in the exciton binding energy~\cite{Ishihara1989, Even2014, Even2015, Yaffe2015}. Moreover, the highly polar lattice, prone to different degrees of local fluctuations~\cite{Leguy2016} highlights the importance of lattice motion and exciton-phonon interactions in establishing electronic correlations~\cite{Hattori:1976wm, Miyata2017, Wu2017, Wright2016}. Particularly, the HOIP lattice is susceptible to dynamic disorder induced by local lattice motion related to vibrations of the organic moiety~\cite{Even2014a}. 
This gives rise to a complex disordered energy landscape, governed by both static and dynamic disorder, that depends on the structure imposed by the organic templating ligand. 
This makes the electronic characteristics of the material sensitive to the organic-inorganic lattice interactions, which have been previously considered in 2D HOIPs~\cite{Cortecchia2017a, Cortecchia2017, Smith2017a, Smith2017, Soe2017}. However, fundamentally important questions remain concerning the role of the dynamic disorder on their ability to sustain strong excitonic multi-particle correlations as seen in other 2D materials. In particular, in this work we ask the following question: given the complex nature of dynamic disorder, are exciton-exciton correlations strong enough to give rise to stable biexcitons at room temperature? 
This is a fundamentally and broadly important question because biexcitons may play a key role in the photophysics relevant to lasing~\cite{Klimov2006}, for example. 
 
The role of biexcitons, bound exciton pairs in two-dimensional HOIPs, has been discussed previously by others~\cite{Ishihara1992,Kondo1998,KATO200315,Fujisawa:2004aa,Elkins2017}. Employing the excitation power dependence of the photoluminescence (PL) spectrum, Ishihara et al.~\cite{Ishihara1992} estimated the biexciton binding energy, the difference between the energy of the bound exciton pair and the energy of two unbound excitons, to be around 50\,meV at low temperature. 
In addition, Kondo~et al. claimed biexciton lasing in a 2D HOIP~\cite{Kondo1998}, and subsequently, Kato~et al.~\cite{KATO200315} measured mutiphoton absorption as well as photoluminescence in a related 2D perovskite, reporting a comparably large binding energy as suggested by Ishihara et al., which was consistent with later measurements on a doped bismuth-doped lead-halide perovskite~\cite{Fujisawa:2004aa}.  More recently, Elkins et al.\ have shown, using two-quantum two-dimensional coherent spectroscopy, that excitons in multi-layered system undergo strong many-body interactions~\cite{Elkins2017}. 

In this work, we address directly excitonic correlations in a prototypical single-layer 2D HOIP. We use temperature as an effective variable to tune the degree of dynamic disorder and to establish its role in the biexciton screening. 
Given the intrinsic complications in the PL lineshape analysis due to temperature dependent linewidth broadening effects, in order to perform a reliable measurement of the biexciton binding energy, we choose to implement two-dimensional coherent excitation spectroscopy~\cite{You2015}.  
By studying a material system consisting of well-defined single-lattice layers that are well separated by templating organic cations, and exploiting the high spectral resolution and selectivity of 2D coherent spectroscopy, we unambiguously determine the biexciton binding energy of \textit{single-lattice-layer} perovskite quantum-well semiconductors, which we find to be comparable to that measured~\cite{You2015} and calculated~\cite{Kylanpaa2015} in strongly excitonic 2D materials such as transition-metal dichalcogenides. We find that biexciton binding energies in these systems are strongly affected by dynamic disorder due to both optical phonons and local lattice vibrations induced by the organic spacer cations. Nonetheless, even in the presence of such strong disorder at room temperature, biexcitons are strikingly stable photoexcitations. This has profound implications for the understanding of electronic correlations in these class of materials, and provide a benchmark for the development of detailed theoretical treatment of their many-body physics.

\section{Results}

\begin{figure}  \centering
    \includegraphics[width=1\textwidth]{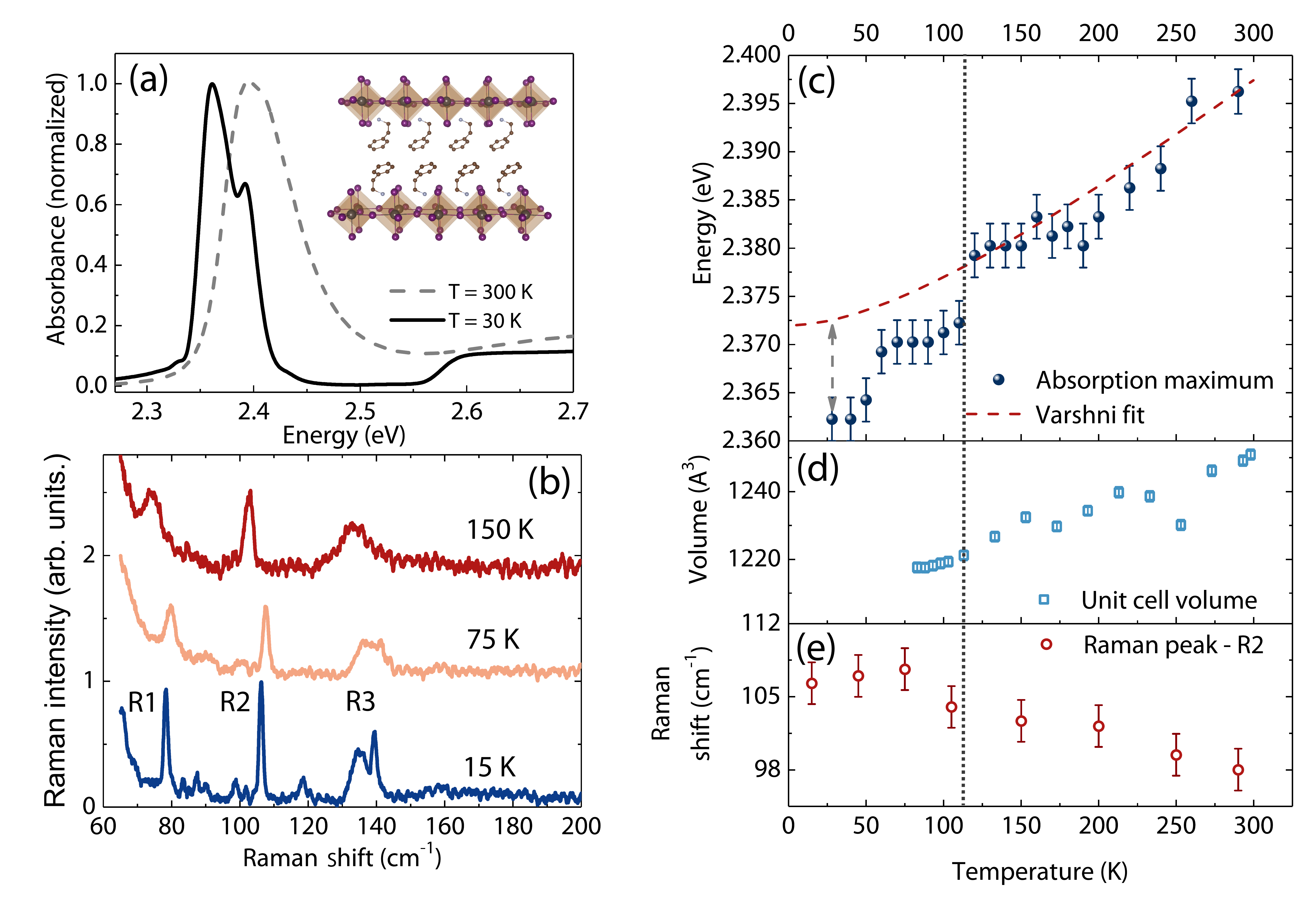}
  \caption{The relationship between optical and structural properties of (PEA)$_2$PbI$_4$. (a) Absorption spectra and structure (inset) of (PEA)$_2$PbI$_4$ monolayer 2D perovskite at low (30\,K) and room temperature (300\,K). (b) Non-resonant Raman spectra of the sample taken at three different temperatures. The main modes identified within this range are labelled as R1, R2, R3 (c) Position of the excitonic peak in the absorption spectrum, plotted as a function of the temperature. Also shown as a dotted line is a Varshini fit of the trend, assuming temperature independence of the exciton binding energy. (d) Unit cell volume estimated from the wide-angle X-ray scattering data shown at various temperatures. (e) Position of the peak R2 in the Raman spectrum plotted at different temperatures. The uncertainty bars represent  instrument resolution.
  \label{absorption_spectrum}}
\end{figure}

We consider polycrystalline films of (PEA)$_2$PbI$_4$ constituting of monolayers of lead iodide octahedra lattices separated by phenylethlyammonium (PEA) cations (see inset of Fig.~\ref{absorption_spectrum}). The wide-angle X-ray scattering (WAXS) data, shown in Figs.~S1--S3 in Supplemental Material~\footnote{See Supplemental Material at [URL will be inserted by publisher] for additional information.}, supports the monolayer architecture. Whilst all the optical characterization presented here has been done on polycrystalline films, the structural data have been extracted from powder X-ray diffraction patterns. Detailed analysis of the WAXS data from films and powdered samples, presented as Supplemental Material~\cite{Note1}, establishes the equivalence of their structural characteristics.

The absorption spectra of the sample under investigation, taken at room temperature and at 30\,K, are shown in Fig.~\ref{absorption_spectrum}(a). Both spectra show a distinct excitonic transition shifted from the band edge by about 200\,meV. At lower temperatures, the exciton peak develops a fine structure, identified initially by Gautron~et al.~\cite{Gauthron:10}, and discussed within the context of a polaron model. A few subsequent works have attempted to explain this phenomenon by invoking phonon vibronic replicas~\cite{Straus2016a,Cortecchia2017a}, although the lack of clear vibrational modes at the frequencies that would be required to account for the spacing of the peaks in the exciton spectrum ($\sim 35$\,meV) makes these assignments inconclusive. We consider that this structured lineshape could arise from other electron-phonon coupling effects that renormalize the excitonic Rydberg energy, thereby producing distinct excitonic states separated by coupling energies. Such effects have been reported in cesium halides~\cite{Moran1965} and in other semiconductors such as GaAs, CuCl, and CdTe~\cite{Geddo1990,Itoh1995}. Alternatively, the spectral structure could arise from Rashba effects. 
In fact, Zhai~et~al.\ report a Rashba splitting of 40\,meV in (PEA)$_2$PbI$_4$~\cite{Zhai2017}, consistent with the energy spacing of the features that we observe in the linear absorption spectrum at low temperature. 
Irrespective of the physical origin of their structure, the spectra confirm the presence of four distinct excitonic transitions separated by about 35--40\,meV. Although they become more evident at low temperature due to reduced thermal disorder, their presence even at room temperatures cannot be discounted, as will be demonstrated by our coherent nonlinear spectroscopy measurements discussed below. 

A closer inspection of the temperature dependence of the exciton absorption spectrum reveals the relationship between crystal and electronic structures. Fig.~\ref{absorption_spectrum}(c) shows the exciton peak absorption energy plotted as a function of temperature (for the complete spectra, refer to Figs.~S4--S5 in Supplemental Material~\cite{Note1}), and we observe a red shift with decreasing temperature. This trend is 
monotonic until about 120\,K and can be explained via lattice contraction effects. The unit cell volume reduces upon lowering the temperature, as shown in Fig.~\ref{absorption_spectrum}(d) (see Fig.~S3 and Fig.~S6 in Supplemental Material~\cite{Note1} for more details), which subsequently reduces the bandgap ($E_G$). Assuming that the exciton binding energy ($E_X$) is temperature independent, the exciton peak energy, $E_0(T) = E_G(T) - E_X$, will follow a similar trend. This dependence can be fitted by an empirical Varshni law~\cite{Varshni1967}, given by $E_0(T) = E_0(0) + \alpha {T^2}/{(T+\beta})$, shown as dashed line in Fig.~\ref{absorption_spectrum}(c). The fit parameters used here are, $E_0(0) = 2.372$\,eV, $\alpha = 1.1 \times 10^{-4}$\,eV\,K$^{-1}$ and $\beta = 151$\,K. Concomitantly, we also observe an increase in the energy of the vibrational modes, as seen in the temperature dependence of the Raman peak $R2$ in Fig.~\ref{absorption_spectrum}(e), which corresponds to an optical longitudinal phonon~\cite{Ni2017}. The temperature dependence of the energies of the other Raman modes is shown in Fig.~S7 in Supplemental Material~\cite{Note1}. 

Below approximately 100--120\,K, we observe a sharp deviation from the behavior described above. The exciton energy now shows abrupt shifts with decreasing temperature over relatively well defined temperatures. The structural parameters extracted from WAXS and Raman spectroscopy also show distinct trends. It must be emphasized that we did not observe a first order phase transition over the measured temperature range by means of WAXS measurements, with the lowest probed temperature of 83\,K. This is also in agreement with earlier works on this material~\cite{Ishihara1996}.  Instead, at this lower temperature range, the contraction of the unit cell with decreasing temperature appears to level off (Fig.~\ref{absorption_spectrum}(d)), although we recognize that the experimentally available temperature range is limited. Below $\sim 100$\,K, we also note that the Raman linewidths (R1 and R2) are weakly temperature dependent down to 15\,K. In conjunction, these observations suggestmpossible stiffening of the crystal lattice as the two inorganic layers come closer, largely due to the steric hindrance between the organic moieties.  

While the Raman shift itself evolves into an anomalous range, where the energy reduces upon lowering the temperature, the most striking evolution at low temperature can be seen in the spectral lineshape, particularly that of the mode labelled $R3$ in Fig.~\ref{absorption_spectrum}(b). We observe a well-defined peak within this vibrational band, which gains intensity at lower temperatures. 
Based on earlier work~\cite{Abid1994, Caretta2014}, we assign this mode to the motion of the lead iodide octahedra induced by the relative motion of the the organic cation. Due to the localized nature of these vibrations and to the large dynamic disorder intrinsic to the perovskite lattice, the observed mode is usually broadened via inhomogeneous effects as can be seen in the 150-K spectrum (Fig.~\ref{absorption_spectrum}(b)). We consider that a sub-ensemble of oscillators perceive similar local environment below 100\,K due to a disorder-order-like transition, as also suggested by the apparent lattice stiffening, leading to the narrowing of the energetic distribution perceived by $R3$. This establishes temperature as an effective means to vary the degree of dynamic disorder within the lattice and thus to investigate the role of the latter in the excitonic correlations. We note that temperature-dependent spectral narrowing of $R1$ and $R2$ modes is weak in this lower-temperature regime, further pointing to the likely lattice stiffening effects.

Such a change in the order parameter of the lattice indeed results in the re-normalization of the exciton binding energy by about 10\,meV. Although this amounts to less than a 5\% correction to the two-particle correlation energy, it nevertheless suggests a finite contribution of the lattice fluctuations to the electronic polarization. There is indeed an observable evolution in the spectral fine structure of the excitonic line in absorption below 100\,K(shown in Fig.~S5 in Supplemental Material~\cite{Note1}) due to non-trivial corrections in exciton-phonon coupling. However, the relatively small change in the binding energy indicates that intrinsic electron-hole interactions are not strongly perturbed by the slow lattice vibrations. 
This may not necessarily be the case for bound multi-excitons, which are expected to have binding energies that are a fraction of the single exciton binding energy as suggested by intensity-dependent photoluminescence measurements~\cite{Ishihara1992}. %
In order to explore this proposition, we require a measurement of the excitation lineshape that permits direct identification of multiquantum resonances, that is, spectral signatures of exciton-exciton correlations, beyond the linear absorption probe discussed thus far, and beyond steady-state photoluminescence, where the biexciton contribution needs to be extracted from within the inhomogeneously broadened exciton spectrum over the \emph{entire} temperature range of interest. We achieve this by performing multi-dimensional coherent spectroscopy at room temperature and at 5\,K, the two experimentally accessible extremes of the range in dynamic disorder of the lattice considered in Fig.~\ref{absorption_spectrum}. The key conclusion that we will draw from this spectroscopy is that the excitonic spectral structure discerned from linear absorption is on top of two-quantum contributions due to biexcitons. We measure directly their binding energy and find that biexcitons are strongly bound at room temperature in spite of the strong dynamic disordered highlighted above. In what follows we first present a general discussion of 2D coherent spectroscopy, and how it enables us to draw these conclusions. 

\begin{figure}  \centering
   \includegraphics[width=\textwidth]{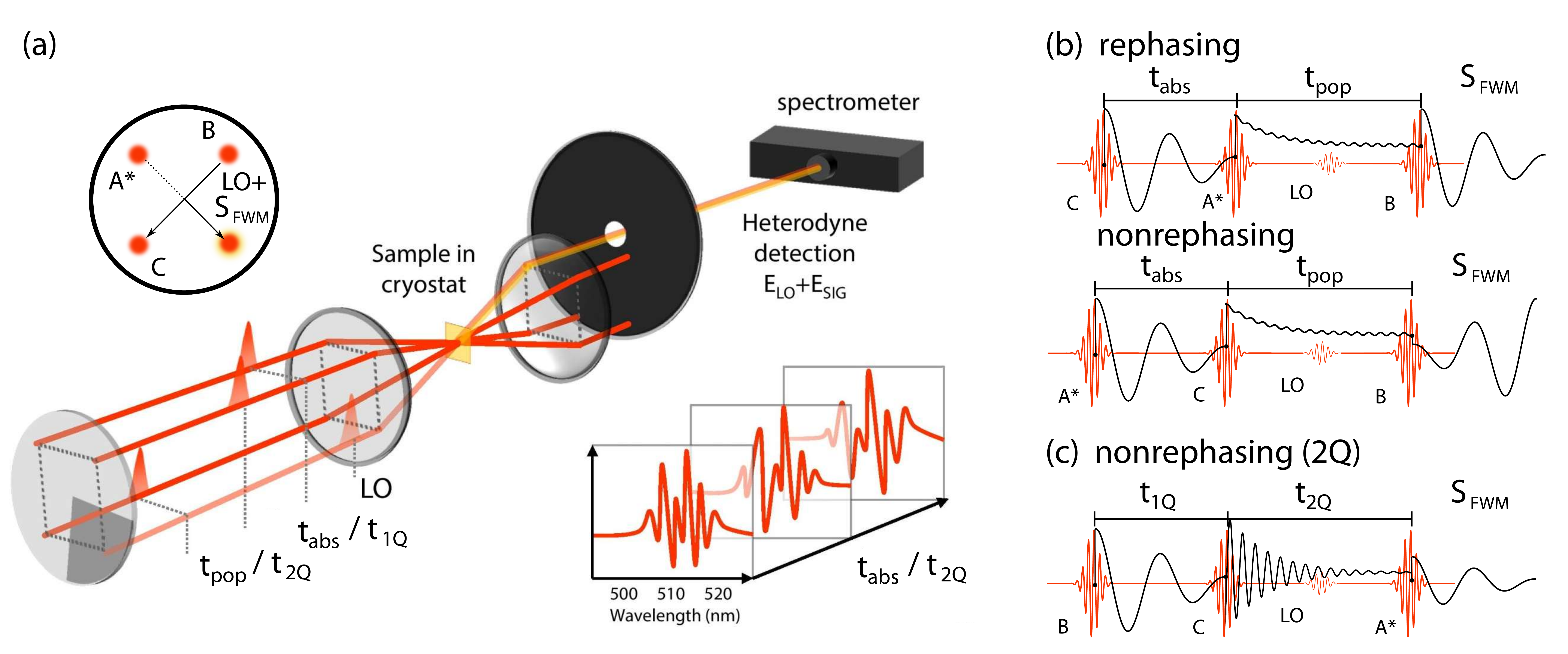}
 \caption{Schematic representation of the two-dimensional coherent spectroscopy experiment implemented in this work. The geometry of the excitation pulse-train beam pattern (red) and the resonant four-wave mixing signal ($S_{\mathrm{FWM}}$, yellow-orange), detected by interference with a local oscillator (LO), is shown in (a). We use a so-called BOXCARS beam geometry, in which three pulse trains (A, B, C) propagating along the corners of a square are focused onto the sample with a common lens, defining incident wavevector $\vec{k}_\mathrm{A}$, $\vec{k}_\mathrm{B}$, and $\vec{k}_\mathrm{C}$. The LO beam, on the fourth apex of the incident beam geometry, co-propagates with $S_{\mathrm{FWM}}$ with wavevector imposed by the chosen phase matching conditions. The spectral interferogram of $S_{\mathrm{FWM}}$ and the LO beam is recorded at every time step. (b) By controlling the order of the pulse sequence with this beam geometry, we measure two distinct $S_{\mathrm{FWM}}$ responses: the \textit{rephasing} signal with wavevector $\vec{k}_\mathrm{B} - \vec{k}_\mathrm{A}  +\vec{k}_\mathrm{C}$ and \textit{nonrephasing} signal  with wavevector $\vec{k}_\mathrm{A} - \vec{k}_\mathrm{B}  +\vec{k}_\mathrm{C}$. The sum of rephasing and nonrephasing spectra produce the total (one-quantum) correlation spectra displayed in Fig.~\ref{1q}. (c) By imposing the depicted pulse sequence, we isolate the two-quantum coherence correlation spectra presented in Fig.~\ref{2q}. 
  \label{setup}}
\end{figure}

Two-dimensional (2D) coherent spectroscopy is a powerful tool to disentangle congested spectral features by measuring the correlations between them~\cite{Fuller:2015xz}. In this spectroscopy, we resolve a nonlinear optical response of the system (coherent radiation from a nonlinear polarization induced in matter by a sequence of three femtosecond pulses) with two correlated energy variables, the ``absorption'' and  ``emission'' energies. The response along the absorption energy variable is extracted from a time-domain coherent excitation spectral measurement using a sequence of the three phase-controlled femtosecond pulses, and that along the emission energy variable is obtained by measuring the resulting coherent emission by means of spectral interferometry with a fourth replica pulse. We represent schematically the geometry of the pulse sequence and resulting signal vectorial direction in Fig.~\ref{setup}. The utility of this family of techniques is that one expects to reproduce the spectral structure observed in the absorption spectrum, manifested by peaks along the diagonal of the 2D spectrum, which expresses optical-transition autocorrelations. On top of this diagonal structure, we can expect off-diagonal cross peaks in the presence of correlations between optical transitions. Furthermore, if a contribution from an excited-state absorption is possible (such as would be expected for a multi-exciton contribution to the absorption spectrum), it would be obscured in a linear measurement or in an incoherent nonlinear measurement such as transient absorption, but it can be uniquely identified in a coherent 2D spectral experiment. In its most common implementation, 2D coherent excitation spectroscopy is achieved by measuring the spectral phase and amplitude of the transient four-wave mixing signal generated by the invoked pulse sequence incident on the sample (see Fig.~\ref{setup}(a)). The delays between the pulses are scanned and the coherent emission spectrum is recorded by spectral interferometry with the ``local oscillator'' laser beam at each time step to generate a time-frequency map of the nonlinear response. The time variable that maps the evolution of the coherence dynamics is then Fourier-transformed to generate a frequency-frequency map.  
The multidimensional spectrometer used in this letter is colloquially referred to as COLBERT (Coherent Optical Laser BEam Recombination Technique), a design of superior phase stability developed by Turner and Nelson~\cite{Turner2016}. More details concerning the implementation used in this work are presented in Appendix D and in Section~S3 in Supplemental Material~\cite{Note1}.

Our experimental configuration allows us to perform two distinct experiments represented schematically in Fig.~\ref{setup}. The first one, the one-quantum (1Q) total correlation measurement, probes transitions between the ground state and the first ladder of excited states, as well as the associated transitions to the next ladder of excited states. It is obtained by summing the real parts of the two-dimensional spectra corresponding to the sequences depicted in Fig.~\ref{setup}(b), the so-called rephasing and nonrephasing spectra. In such spectra, coherence or population transfer between spectral features and excited-state absorption (ESA) manifests as negative and positive features respectively~\cite{Hamm2011}. These two signatures are often spectrally degenerate and give rise to complex lineshapes in a 1Q spectrum. 
The second experimental scheme involves two-quantum (2Q) nonrephasing measurements, and exclusively probes the direct two-quantum transitions to the second excited-state manifold, making it a more selective measurement of higher-lying states. One performs such a measurement by using the pulse sequence depicted in Fig.~\ref{setup}(c). It is thus a very useful tool to interpret 1Q spectra ESA features. Due to its high selectivity, ease of interpretation and low time-averaged fluences, these measurements have been used as elegant probes of many-body interactions in various systems such as GaAs quantum wells~\cite{Stone2009a}, semiconductor quantum dots~\cite{Moody2014}, monolayers of transition-metal dichalcogenides~\cite{Hao2017}, and multi-layered two-dimensional perovskites~\cite{Elkins2017}, to name a few closely related examples.

\begin{figure}  \centering
    \includegraphics[width=\textwidth]{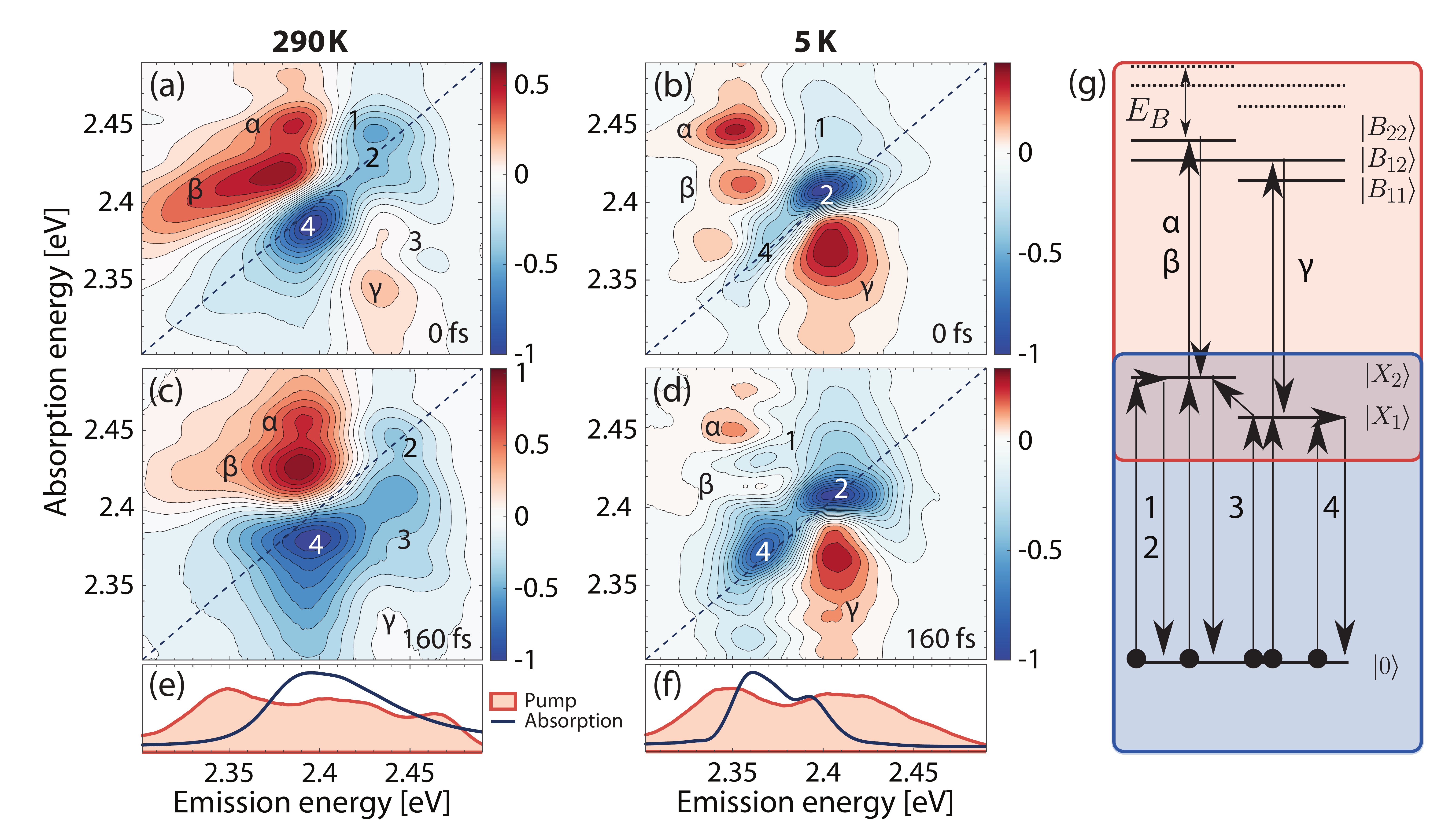}
  \caption{Total-correlation 2D coherent excitation measurements of (PEA)$_2$PbI$_4$.1Q total correlation spectra of (PEA)$_2$PbI$_4$ monolayer 2D perovskite at 5\,K (b,d) and room temperature (a,c) for population times of 0\,fs (a,b) and 160\,fs (c,d). For comparison, the corresponding absorption spectra and pump laser spectra used during the experiments are shown in (e) and (f). An illustration depicting some of the coherent processes involved in these features is presented in (g). To avoid a crowded diagram, transitions within the same exciton manifold (such as $\alpha$ and $\beta$) have been drawn as degenerate. 
  \label{1q}}
\end{figure}

The 1Q total correlation spectra for (PEA)$_2$PbI$_4$ are shown in Fig.~\ref{1q} for two different population times (0 and 160\,fs) at both room temperature ((a) and (c)) and 5\,K ((b) and (d)). It must be noted that the amplitudes of the observable features are weighted by the product of the pump and absorption spectra, which are also shown in Figs.~\ref{1q}(e) and (f). The pump spectrum does not extend into the free-carrier absorption band, allowing us to uniquely probe excitons and their multi-body counterparts. We also present real and imaginary parts of both rephasing and nonrephasing components for both population times in Figs.~S9 and S10, along with a movie of these spectral evolutions, in Supplemental Material~\cite{Note1}. 

All spectra harbor negative diagonal (such as peaks 2 and 4) and off-diagonal (such peaks as 1 and 3) features corresponding to the primary transitions from the ground to the exciton states. Strikingly, large positive features are also present, labelled as $\alpha$, $\beta$ and $\gamma$. These indicate excited-state absorption from single-quantum states to higher-lying ones, suggesting the existence of a non-trivial excited-state manifold (see Fig.~\ref{1q}(g) for a summary of the excitation pathways).  These features lay atop negative-going cross peaks, as indicated by the population time dependence of the diagonal features (see Supplemental Material~\cite{Note1}), obscuring their true lineshapes. Nevertheless, it is possible to estimate the energy of the higher-lying states responsible for these features by comparing their position along the emission axis in the total correlation maps to upper cross-peak labelled 1 and the diagonal feature labelled 2, respectively~\cite{Hamm2011}. For the features in the upper left corners of the spectra (labelled as $\alpha$ and $\beta$), this implies the existence of a state ($42\pm2$) and ($55\pm2$)\,meV below twice the energy of the 1Q state for room temperature and 5\,K, respectively. Supposing that this higher-lying state consists of a bound biexciton, which is a correlated two-electron, two-hole quasi-particle, this energy difference corresponds to our first estimate of its binding energy ($E_B$). The positive feature labelled $\gamma$ can also be caused by ESA into a bound biexciton of two different excitons. In this case, an estimation of the biexciton binding energy is challenging because of the higher proximity and thus higher degree of interference with its corresponding negative feature.

The spectral evolution of the 1Q signal at 5\,K~\cite{Note1} suggests electronic relaxation within the exciton manifold. At initial times, the diagonal response is mainly concentrated at higher energies  and within $< 100$\,fs, the spectral weight is transferred to the lower-energy resonance via population relaxation. Concurrently, cross peaks associated with the correlation of the lower-energy state with the higher-lying ones gain in intensity, enough to partially cover the positive ESA features.

At room temperature, we do not observe such dominant \textit{relaxation} dynamics apart from a monotonic decay over the entire spectrum (see also Fig.~S11 in Supplemental Material~\cite{Note1}). This can be attributed to the lack of spectral structure, even in linear absorption, due to the disorder-induced broadening of the excitonic transitions. We presume that thermal disorder stabilizes a distribution of the photoexcited population within the excited-state manifold, with no substantial relaxation at least within the probed temporal window. On the other hand, we observe that the initial diagonally elongated spectral lineshape broadens along the anti-diagonal axis within tens of femtoseconds. Such behavior is indicative of spectral diffusion due to the energetic re-distribution between the inhomegeneously broadened oscillators via incoherent interactions with the surrounding environnement~\cite{Kozinski2007}.  
The observed ultrafast timescales, though compelling and probably analogous to the ultrafast lattice re-organization dynamics observed in three-dimensional perovskites~\cite{Batignani2017}, require more detailed analysis. 
Nevertheless, cross-peaks arising from these spectral dynamics and from correlations add ambiguity to the estimated \textbf{$E_B$} from the 1Q measurement alone. To alleviate this, we exclusively probe the higher-lying excited-state manifold by performing a two-quantum (2Q) measurement.

\begin{figure}  \centering
    \includegraphics[width=0.6\textwidth]{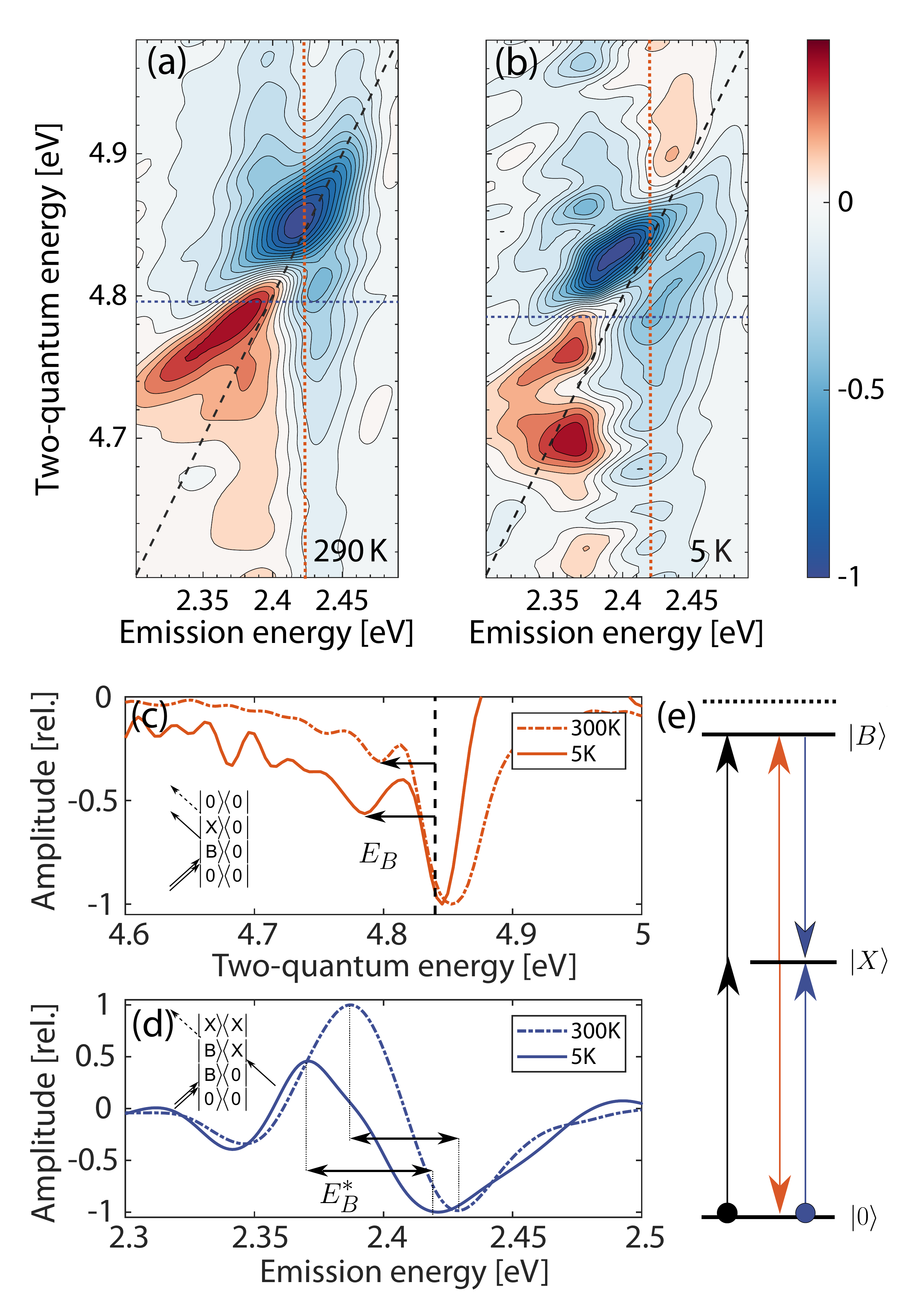}
  \caption{Two-quantum nonrephasing 2D coherent excitation measurements of (PEA)$_2$PbI$_4$. Real part of the 2Q non-rephasing spectra of polycrystalline monolayer (PEA)$_2$PbI$_4$ at room (a) and low temperature (b) for a 1Q waiting time of 20\,fs. The dashed black line follows two-quantum energies at twice the emission energy. Cuts along the vertical (c) and horizontal (d) axes were taken along the dashed lines of corresponding color in (a) and (b). The horizontal cuts were taken to coincide with the two-quantum energy at which a biexcitonic feature resides on the vertical axis. Estimates of $E_B$ and $E^*_B$  were obtained from apparent shifts from the diagonal line or from bound biexcion coherences respectively (see text). Insets of (c) and (d) show the Feynmann pathways responsible for these shifts, with $|X\rangle$ the exciton and $|B\rangle$ the biexciton states, respectively. The dashed line in (c) indicates the crossing point between the cut axis and the diagonal. (e) An representation of the coherent pathways that lead to biexciton features below the $\omega_{2Q}=2\omega_{emit}$ diagonal. 
  \label{2q}}
\end{figure}

The real part of the nonrephasing 2Q spectra of polycrystalline monolayer (PEA)$_2$PbI$_4$, both at 5\,K and room temperature, are shown in Fig.~\ref{2q} alongside a schematic of the energy levels of the excitons $|X\rangle$, biexcitons $|B\rangle$ and unbound exciton pairs (dashed line) involved in the generation of the 2Q signal. The delay between the first and second pulses is set to 20\,fs, chosen to be their temporal full width at half-maximum, to minimize contributions from undesired non-resonant four-wave mixing while still maintaining a high signal-to-noise ratio. 
In both spectra, a strong dispersive lineshape is present on the two-diagonal ($\omega_{2Q}=2\omega_{emit}$) axis indicating unbound but correlated exciton pairs by  many-body interactions such as excitation induced dephasing and excitation induced shift~\cite{Karaiskaj2010}. 
We emphasize that any spectral feature on the two-diagonal axis reflects correlations between two excitons that include neither attractive nor repulsive contributions. 
 Negative peaks can also be observed above and below these dispersive features, and these signatures do correspond to bound exciton pairs of different and similar species respectively~\cite{Turner2010}. The negative features slightly below the diagonal are due to oscillations of coherences between a 2Q state and the ground state. From these features' vertical shift below the two-diagonal, the biexciton binding energy $E_B$ can be extracted yielding ($44\pm5$) and ($55\pm5$)\,meV for room- and low-temperature measurements, respectively. By comparing this energy with that previously obtained from ESA features in 1Q total correlation spectra, we conclude that these previous features indeed arise from ESA into a bound-exciton-pair state. This consists of the first \emph{direct} measurement of biexcitons in \textit{monolayered} (i.e.\ the most quantum-well-like) two-dimensional perovskites.

Theory also predicts the existence of a second positive feature redshifted by the biexciton binding energy along the emission axis (see inset of Fig.~\ref{2q}(d) and Fig.~\ref{2q}(e)). This peak arises from the oscillations of coherences between a 2Q state and the single-exciton state~\cite{Yang2008}. Such a peak can be observed atop the aforementioned dispersive features and yields another estimate of $E^*_B$, the exciton binding energy. Taking an horizontal cut at the two-quantum energy where the minimum of the biexcitonic peak was previously observed highlights the presence of this feature (see Fig.~\ref{2q}(d)). From the shift along the emission axis between the two extrema of this cut, we extract a value of $E^*_B$ of ($42\pm5$) and ($50\pm5$)\,meV for room- and low-temperature measurements, respectively, corroborating the values of $E_B$ discussed above.

We note that biexciton binding energies are often estimated by PL measurements as previously done by Ishihara et al.~\cite{Ishihara1989}. In fact, we do observe an emission peak which could be of biexcitonic origin in PL (Fig.~S8 in Supplemental Material~\cite{Note1}) and is red-shifted by about 40 \,meV from the primary excitonic emission peak. However, the energetic position of the exciton in emission is contaminated by self-absorption effects due to small Stokes shift and overlapping spectral contributions from the exciton finestructure, leading to an under-estimation of $E_B$. Moreover, local heating effects induced by the high average pump powers required make estimates from PL intensity measurements unreliable.
It has also been shown that a non-linear increase of the PL with pump fluence can also arise from defect-related effects adding to the ambiguity in the assignement based solely on intensity-dependent PL mesurements~\cite{SrimathKandada2016}. Furthermore, at room temperature, the PL spectrum is featureless, making it impossible to even identify biexcitonic signatures over this higher temperature range. This highlights the advantage of using multi-dimensional spectroscopy to identify biexcitons and to measure their binding energy, as we have demonstrated here.

\section{Discussion}

In summary, we provide the first direct observation of biexcitons in monolayered (PEA)$_2$PbI$_4$, using both 1Q and 2Q coherent multidimensional spectroscopy, and we have established that this is so in the presence of strong dynamic energetic disorder at toom temperature. The inferred binding energies are in agreement with the previous report of Ishihara et al.~\cite{Ishihara1992}. and approach those measured in monolayer transition metal dichalcogenides~\cite{Hao2017,You2015}, another bidimensional system of great fundamental and technological interest. We consider that it is highly significant that biexcitons are as stable in 2D perovskites, which are subject to strong dynamic disorder due to the hybrid organic-inorganic nature of the lattice, and to its ionic nature, as in atomically single-layer, purely covalent 2D semiconductors. We also provide a temperature-dependent measurement of biexciton binding energy using two-quantum, two-dimensional spectroscopy, pointing to a similar effect of lattice temperature to that observed for the exciton binding energy. The large change in biexciton binding energy with temperature, as a fraction of the total binding energy, suggests important contributions of the lattice to the permittivity function in two-dimensional perovskites. 

Apart from giving an experimental benchmark for multi-body correlations in these materials, this work highlights the importance of the lattice degrees of freedom in Coulomb screening effects. The contribution of polar lattice vibrations has been successfully unravelled in the case of excitons in polar semiconductors by Kane~\cite{Kane1978} and Pollmann-Buttner~\cite{Pollmann1977} via consideration of an effective permittivity function composed of static and optical frequency dielectric responses. This has been effectively extended even to the case of three-dimensional hybrid perovskites, where the motion of the organic cation has been shown, both theoretically~\cite{Even2014a} and experimentally~\cite{SrimathKandada2016d}, to screen electron-hole correlations. Given the smaller exciton binding energy in bulk perovskites, these minor contributions, though present, are considered to have no substantial effect on the excitation dynamics, especially at low solar densities, though their role at high excitation densities cannot be ruled out. 

Intriguingly, two-dimensional perovskites present a contrasting scenario with respect to their 3D counterparts. Dielectric confinement assures an extremely large exciton binding energy accompanied by a small Bohr radius, such that exciton characteristics remain relatively insensitive to lattice fluctuations due to their localized nature. This can be seen in the very modest correction to the exciton binding energy with decreasing temperature, along with the very similar spectral structure evident in both linear and non-linear (1Q) spectra, upon transitioning from a disordered (RT) to ordered (5\,K) crystal phase. The temperature dependence of the biexciton binding energy, on the other hand, makes dynamic disorder an extremely pertinent parameter in many-body correlations and thus assumes relevance for high-density ($> 10^{18}$\,cm$^{-3}$) dynamics. We interpret this as a consequence of their more delocalized nature implied by the lower biexciton binding energy with respect to the single-exciton binding energy, which makes them more sensitive to dynamic disorder induced by localized lattice fluctuations.

Of all semiconductor systems probed so far, biexcitons in 2D perovskites are amongst the most strongly bound~\cite{Kylanpaa2015}. This indicates that at room temperature and sufficiently high excitation densities, they will be the dominant photoexcitation, yielding an important non-radiative channel for exciton population. An Auger-like channel is expected to play a major role in the performance of lasing devices by increasing the lasing thresholds~\cite{Klimov2006}. The highlighted effects of dynamic disorder must thus be accounted for in the optimization of these promising materials for optoelectronic applications. In the case that biexcitons are the emissive species used for the lasing action as suggested by Kondo et al.\ with 2D perovskites~\cite{Kondo1998}, crystal lattices that can house stable biexcitons should be appropriately designed and optimized.

\section{Conclusions and Outlook}

We conclude that in a model metal-halide \emph{single-layer} two-dimensional hybrid semiconductor, biexcitons are primary photoexcitations at sufficiently high density ($\gtrsim 10^{18}$\,cm$^{-3}$) in spite of a highly complex disordered energy landscape. By means of temperature-dependent absorption, wide-angle X-ray scattering, and Raman spectroscopies, we have associated contributions of lattice motion to the dynamic disorder that renormalizes exciton energies. These lattice dynamics affect substantially biexciton binding energies such that at 5\,K these are 25\% higher than at ambient temperature, read directly by means of two-dimensional coherent excitation spectroscopy. Nevertheless, given our measurement of a binding energy of 44\,meV at room temperature, only $\sim10$\% of biexcitons would dissociate at steady state under ambient conditions. This underlines the importance of studying multi-excitonic structure in these materials as a function of chemical and crystalline structure. We consider it an opportunity to extend these ultrafast spectroscopic studies as a function of lead halide octahedral distortion induced by the organic templating cation~\cite{Cortecchia2017a}. We suggest that control of the details of dynamic disorder can be achieved by means of the nature of the organic moiety. This will provide a rich materials parameters space to explore many body correlations beyond atomically-single-layer semiconductors such as transition-metal dichalchogenides.

Multi-particle correlations are at the heart of the quantum phase transitions --- exciton-Mott transitions and Bose-Einstein condensation~\cite{Carusotto2013}, which are the primary mechanisms leading to photonic and polaritonic lasing, respectively, in semiconductors. Conceptualizing the effect of lattice fluctuations on the co-operative behavior of the coherent electronic excitations in such phases is thus not only of technological significance, but also a new frontier for semiconductor physics of highly disordered yet strongly excitonic semiconductors. 

\begin{acknowledgments}
A.R.S.K. acknowledges funding from EU Horizon 2020 via Marie Sklodowska Curie Fellowship (Global) (Project No. 705874). SN and AP acknowledge funding from  EU Horizon 2020 Research and Innovation Program under grant agreement no.\ 643238 (SYNCHRONICS). FT acknowledges a Doctoral Postgraduate Scholarship from the Natural Sciences and Engineering Research Council of Canada and Fond Qu\'eb\'ecois pour la Recherche: Nature et Technologies. CS (Silva) acknowledges support from the School of Chemistry and Biochemistry and the College of Science of Georgia Institute of Technology and the Canadian Foundation for Innovation. RL acknowledges funding from the Natural Science and Engineering Research Council of Canada. CS (Soci) and DC acknowledge support from the National Research Foundation of Singapore (NRF-CRP14-2014-03). 
\end{acknowledgments}

\section*{Appendix A: Sample Preparation}
For the preparation of (PEA)$_2$PbI$_4$ thin films, the precursor solution (0.1\,M) of (PEA)$_2$PbI$_4$ was prepared by mixing (PEA)I with PbI$_2$ in 1:1 ratio in DMSO. For example, 24.9\,mg of (PEA)I and 23\,mg of PbI$_2$ were dissolved in 500\,$\mu$L of DMSO. The solution was heated at 100$^o$C for 1 hour and then spin-coated on glass substrates at 4000 rpm for 30\,s.  Finally, the samples were annealed at 100$^o$C for 15 minutes to obtain the formation of orange films.
For the preparation of (PEA)$_2$PbI$_4$ crystals, PbO (223.2\,mg) was dissolved in 2\,mL of aqueous HI solution together with 170\,$\mu$L of 50\% aqueous H$_3$PO$_2$. In a different vial, 92.4\,$\mu$L of phenethylamine were added to 1\,mL of HI 57\% wt, resulting in the formation of a white precipitate which is quickly re-dissolved under heating. This solution was then added to the PbO solution, and the mixture was stirred at 150$^o$C on a hotplate. The stirring was stopped after 10 minutes, leaving the solution to cool down at room temperature. After 24 hours, the resulting orange crystals were collected by filtration and dried at 100$^o$C under vacuum. Perovskite powders for WAXS characterization were obtained by gently grinding the crystals in a mortar.

\section*{Appendix B: Absorption and Raman spectroscopies}
Temperature-dependent absorption measurements were carried out using a commercial Perkin-Elmer UV/Vis spectrophotometer. The sample was kept in a continuous flow static-exchange gas cryostat (Oxford Instruments Optistat CF). Measurements were taken in steps during the heating up cycle, after going to liquid helium temperatures. The non-resonant Raman measurements at 718\,nm excitation wavelength were taken  in  a  near  backscattering  configuration  with  a  continuous-wave Ti:Sapphire  laser  (Spectra Physics, Matisse TS) using 100\,mW excitation power. The beam was focussed on the sample by a cylindrical lens, resulting in a  spot  size  of  about  1\,cm $\times$ 50\,$\mu$m .  The spectrum was detected by a liquid nitrogen cooled Princeton Instruments CCD in conjunction with a Jobin-Yvon U1000 double spectrometer.

\section*{Appendix C: Temperature-dependent X-ray diffraction measurements. }
Thin film X-Ray diffraction was performed in a BRUKER D8 ADVANCE with Bragg Brentano geometry, $Cu K\alpha$ radiation ($\lambda = 1.54056$ {\AA}), step increment of 0.02$^o$, and 1 s of acquisition time. The SAXS/WAXS system Nano-inXider from Xenocs (power range 50\,kV, 0.6\,mA) equipped with Linkam HFSX350 temperature stage was used for temperature-dependent measurements of perovskite powders. The measurements were performed with beam size of 200\,$\mu$m, flux 4\,Mph/s, cooling-heating rate of 3$^o$C, 1\,min stabilization time and 5\,min exposure time. The software TOPAS 3.0 was used to perform Pawley fit by fundamental parameters approach. The starting lattice parameters were taken from the reference structure reported by J. Calabrese et al.~\cite{Calabrese1991}. Peak profile and background were fit, respectively, with a TCHZ Pseudo-Voigt function and a Chebichev polynomial of fifth order with $1/x$ function. The zero error, scale factor, linear absorption coefficient, and lattice parameters were refined during the fitting. 

\section*{Appendix D: Multidimensional spectroscopy}
The multidimensional spectrometer used in this work consists of COLBERT, a design of superior phase-stability developed by Turner and Nelson~\cite{Turner2016}. Such phase stability is passively achieved by delaying the pulses using phase-shaping only and propagating the beams  through a common set of optics. The spectrometer uses pulses generated by a home-built two-pass non-collinear optical parametric amplifier (NOPA) pumped by the output of a 1-kHz Ti:Sapphire regenerative amplifier system (Coherent Astrella). Before entering COLBERT, the NOPA's output is spatially filtered using a pinhole to obtain a clean Gaussian profile. COLBERT's pulse-shapper is also used to compress the pulses near the transform limit using chirp scan~\cite{VincentL2013} and MIIPS~\cite{Xu2006a}. For the dataset presented here, this resulted in pulses of 21-fs temporal full-width at half-maximum (FWHM) as measured by second harmonic generation collinear frequency resolved optical gating (SHG-CFROG)~\cite{Amat-Roldan2004a}. More details concerning the pulse temporal characterization and compression can be found in Section~S5 in Supplemental Material~\cite{Note1}.

Four beams with 3-mm diameter were focused on the sample  using a 20-cm achromatic lens to generate the non-linear signal. The beams were generated by sending the input beam through a diffractive optical element (DOE) designed such that the transverse wavevectors of each beams resides in the opposite corners of a square. Each pulse carried 15\,nJ amounting to a fluence of about 20\,$\mu$J/cm$^2$ except the beam used for heterodyne detection which was attenuated a thousandfold to prevent any non-linear interaction with the sample. The sample was held in a cold-finger closed-cycle cryostat (Montana Instruments Cryostation) with the active layer in firm contact with the cold finger.

\section*{Appendix E: Author Contributions}
FT and SN collected the 2D spectroscopy data and carried out their analysis, supervised by ARSK and CS (Silva). SN and VAD measured the Raman spectra, supervised by RL. SN measured the absorption data, supervised by AP and ARSK. TS and LYM carried out the temperature-dependent X-ray scattering measurements. DC carried out the synthesis of the materials, supervised by CS (Soci). FT, SN, ARSK, and CS (Silva) led the conceptual development of the project. All authors contributed to the redaction of the manuscript. ARSK and CS (Silva) are to be considered co-principal investigators, and FT and SN are to be considered first co-authors.


%

\newpage


\section*{Supplementary material}

# Supplementary Information for
# Stable biexcitons in two-dimensional metal-halide perovskites with strong dynamic lattice disorder

## S1 Structural characterization

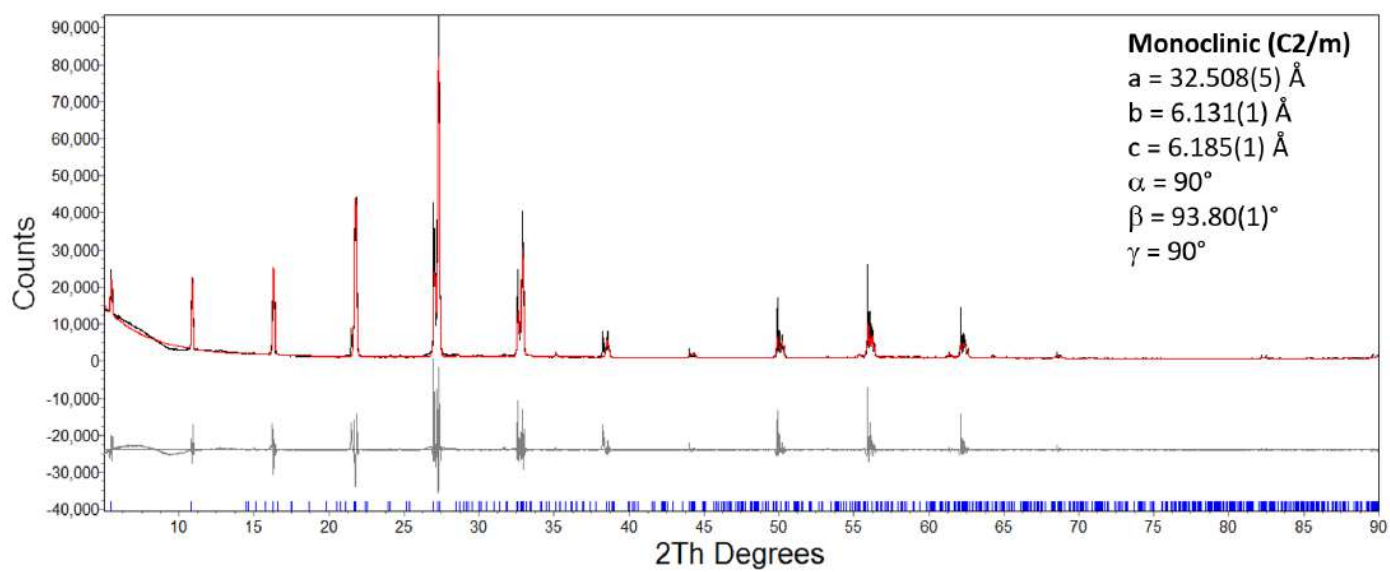

**Figure S1:** *X-ray diffraction pattern of $(PEA)_2PbI_4$ crystals before grinding and corresponding Pawley fit: experimental (blue) and calculated (red) diffraction pattern. The grey line is the difference between the observed and experimental pattern, while the blue marks indicate the reflections expected for the corresponding perovskite phase*

Fig.S1 shows the XRD pattern of the $(PEA)_2PbI_4$ single crystals and corresponding Pawley fit, before grinding. The diffraction pattern is fully consistend with the expected monoclinic structure (C2/m). The *h00* reflections are enhanced due to the strong preferential orientation of the crystals in this direction. For the temperature dependent measurements, powedered crystals are used, Fig.S2(A) and FigS2(B) show the Pawley fit of the wide-angle X-ray scattering (WAXS) signal measured on the powders at room temperature and at 83 K, respectively. Fig.S2(C) shows the Pawley fit of the XRD pattern of thin films of $(PEA)_2PbI_4$ using the lattice parameters extracted from powder diffractograms at 298 K and establishes the equivalence of XRD analysis of powdered crystals and thin films. Powder WAXS patterns for the entire temperature range is plotted in Fig.S3(A) and the comparison between the RT and 83 K is explicitly shownn in Fig.S3(B). We did not observe any phase transition in the entire temperature range. The change in the diffraction pattern is consistent with the gradual shrinkage in the unit cell volume reducing the temperature, but the material maintains the same phase.

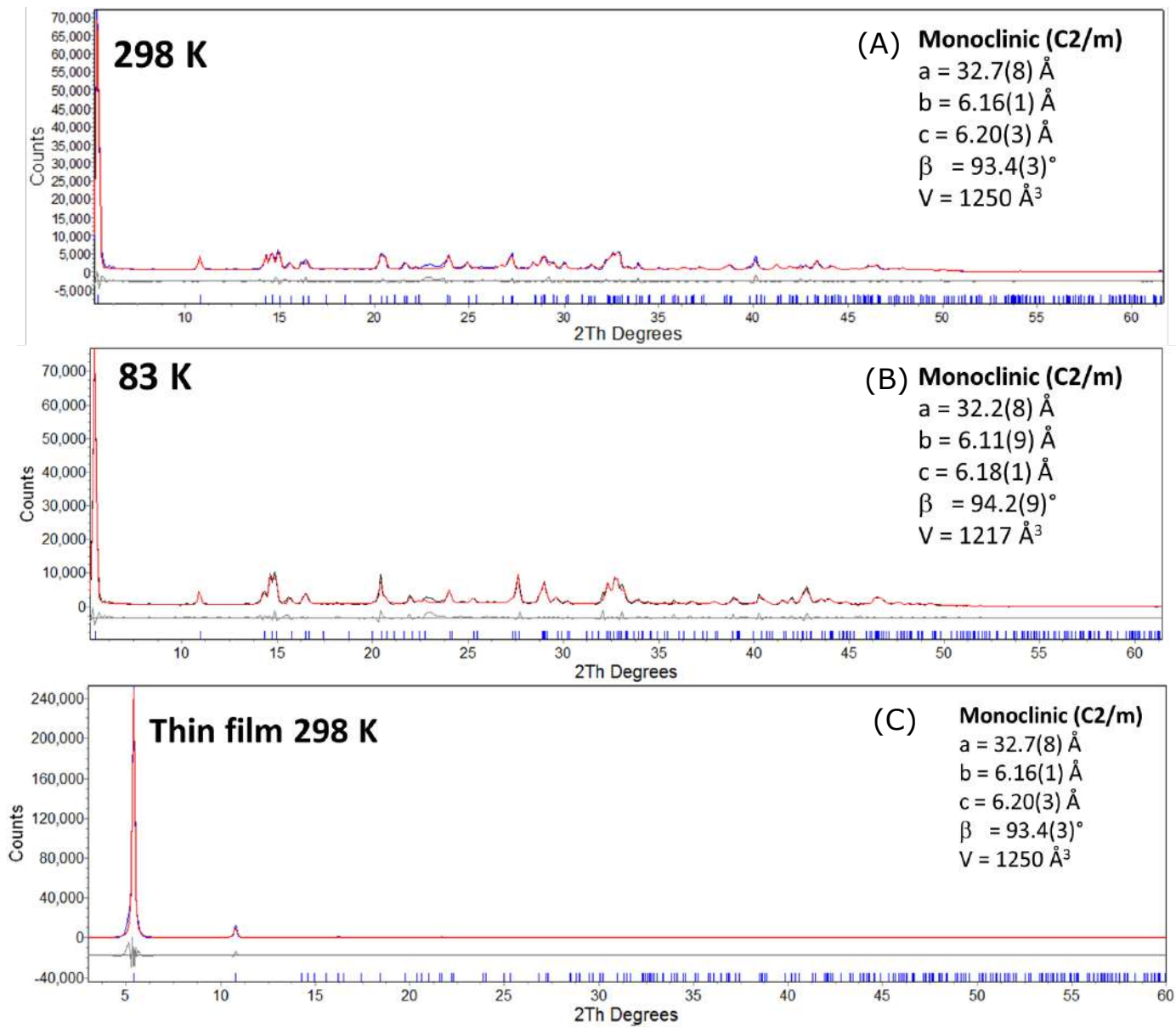

**Figure S2:** *Pawley fit of WAXS signal measured on $(PEA)_2PbI_4$ powders from ground crystals and at (A) 298 K and (B) 83 K. (C) Pawley fit of the signal from the thin film. The film is strongly oriented towards the 200 direction.*

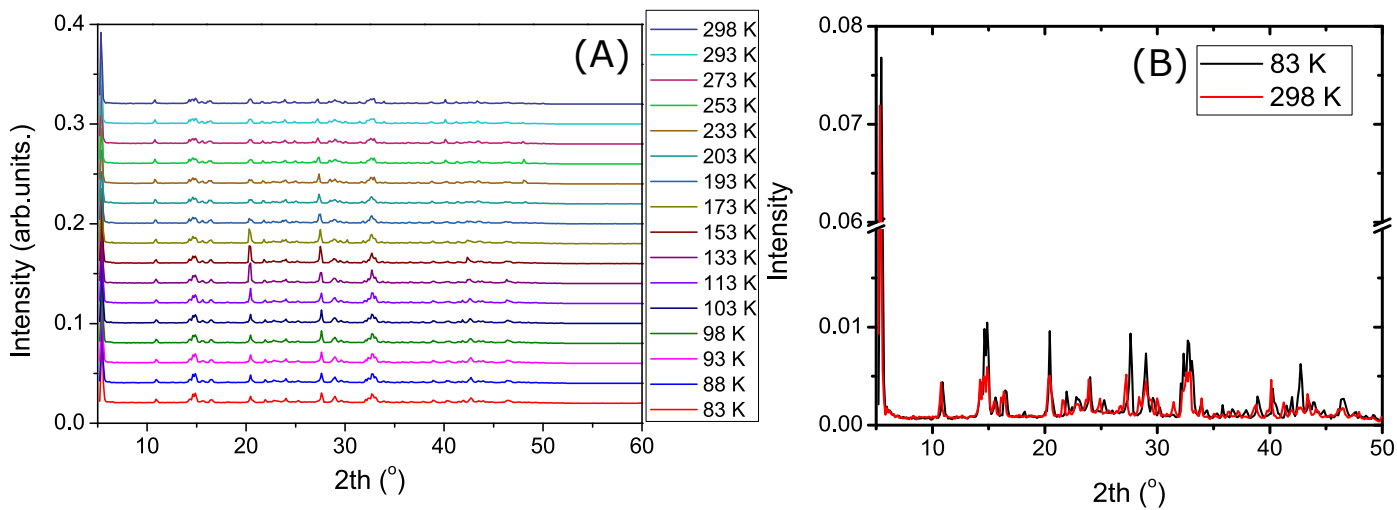

**Figure S3:** *(A) Powder XRD spectra at different temperatures. (B) Comparison between the spectra at 298 K and 83 K*

# S2 Linear absorption

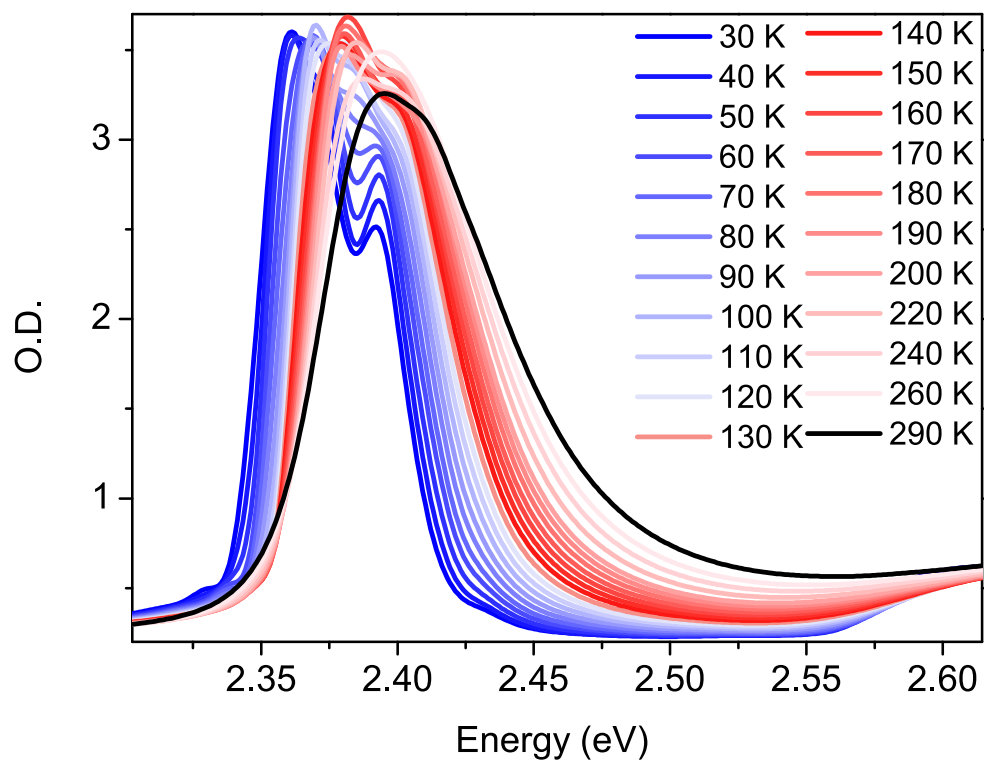

**Figure S4:** *Absorption spectra of the sample taken at different temperatures*

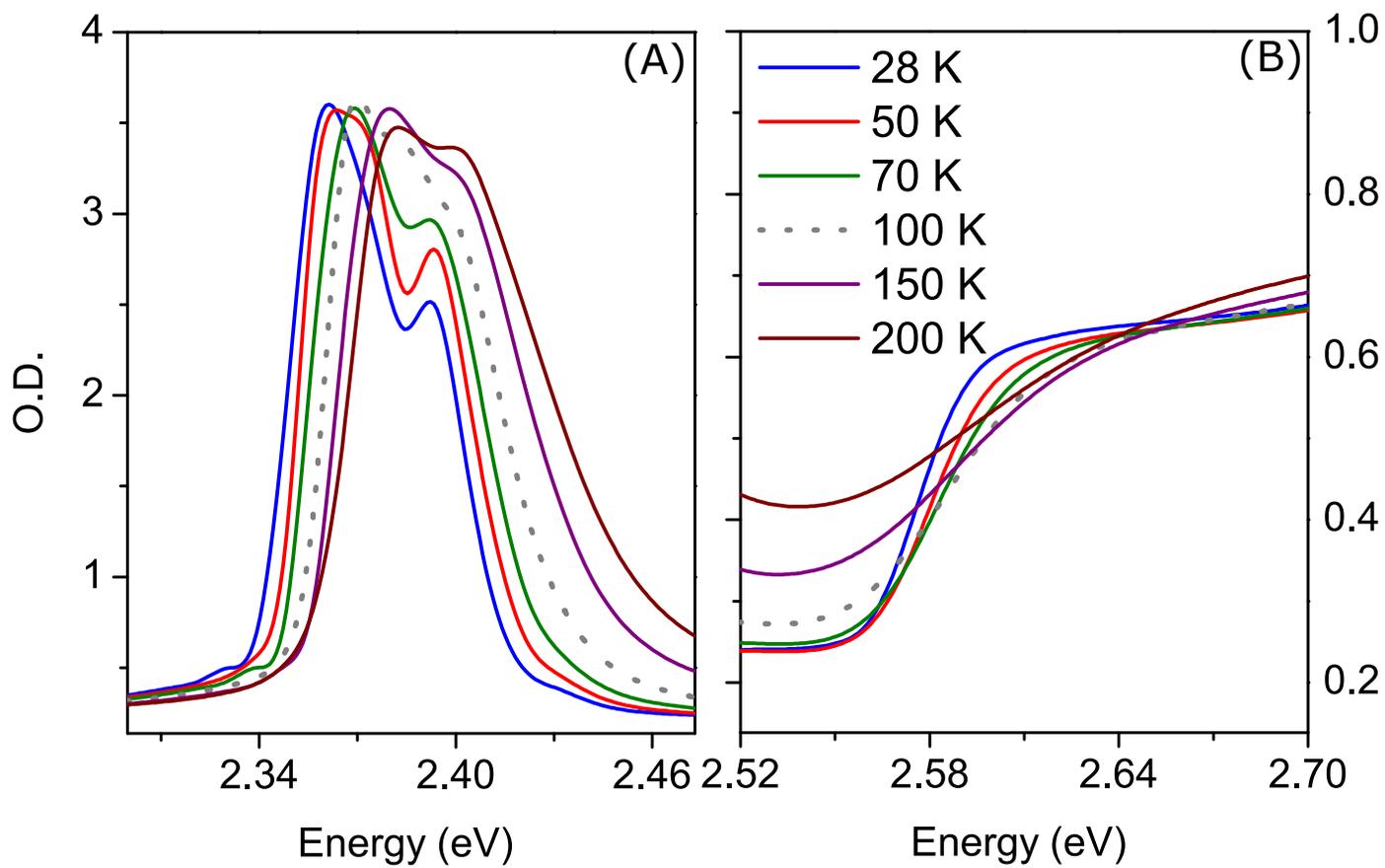

**Figure S5:** *Absorption spectra of the sample taken at different temperatures*

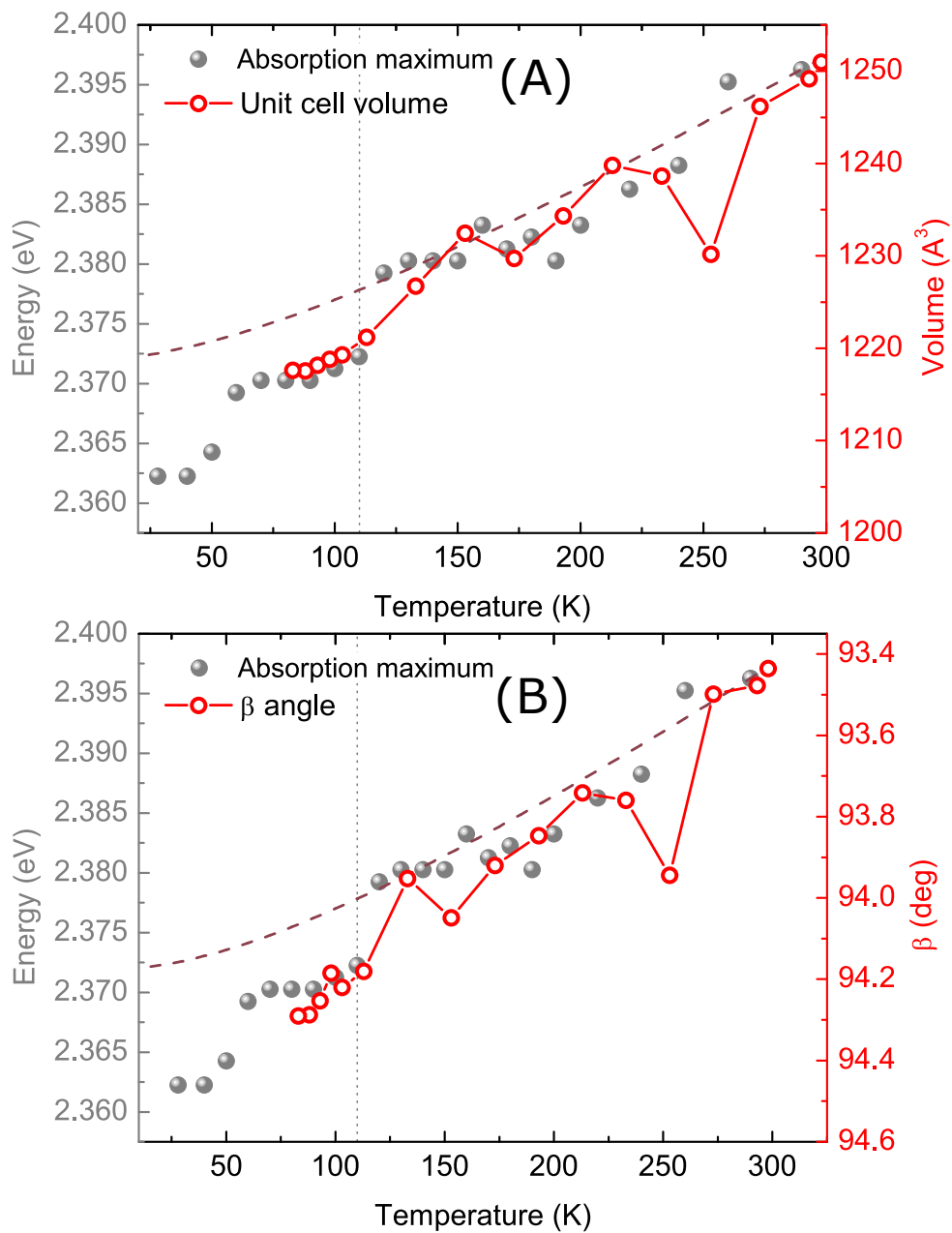

**Figure S6:** *Absorption peak versus structural parameters from XRD - Temperature dependence*

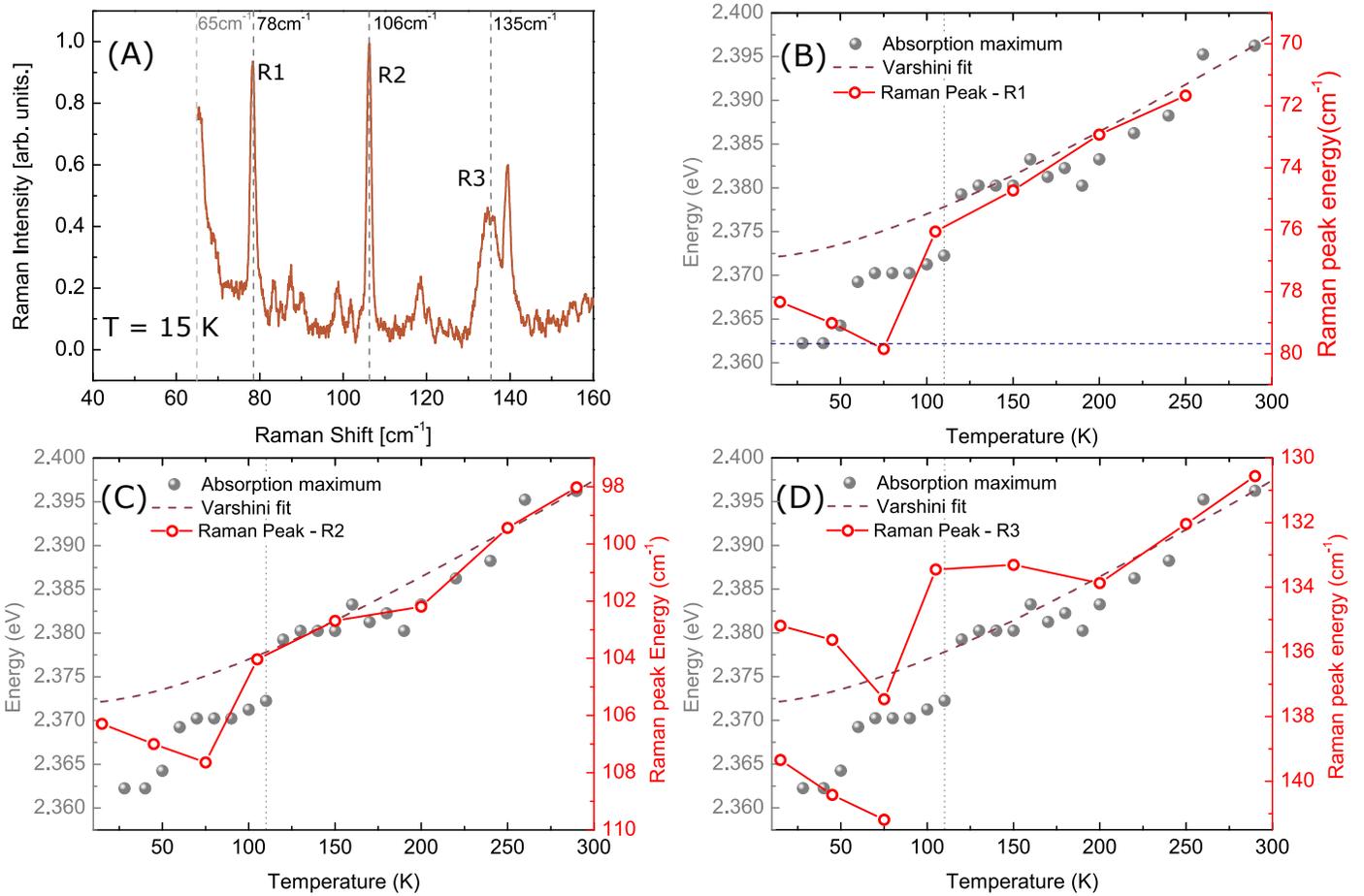

**Figure S7:** *Absorption peak versus Raman modes - Temperature dependence*
6

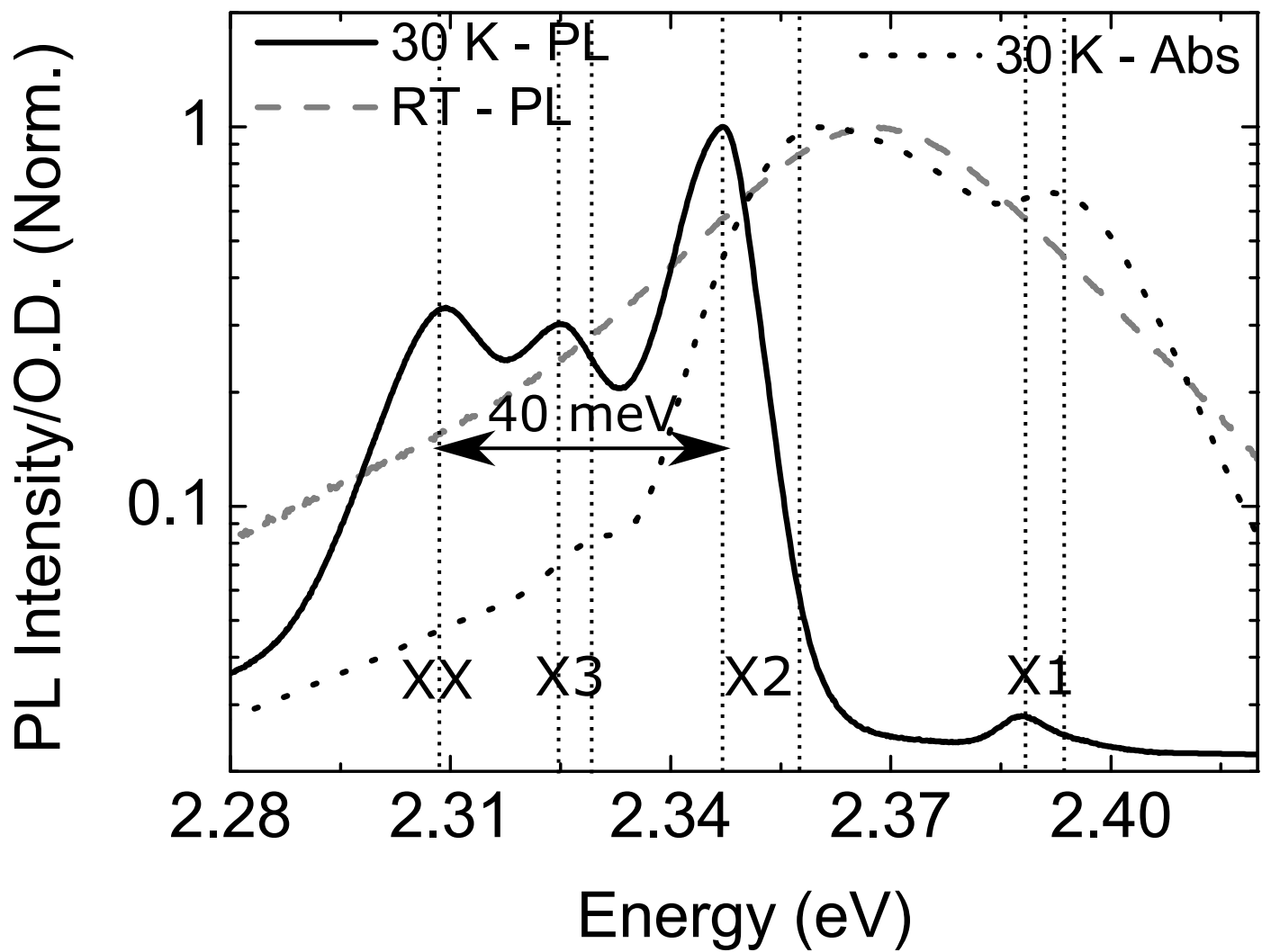

**Figure S8:** *PL - Temperature dependence*

# S3 Rephasing and non-rephasing 1Q spectra

Typically, 3rd order two-dimensional spectra such as the ones presented in this letter are obtained by sending a sequence of 3 pulses on the sample and collecting the resulting four-wave mixing (FWM) signal by interference with a slightly delayed fourth beam called a local oscillator (LO). This allows not only the simultaneous measurement of the spectral amplitude and phase of the signal, but also allows the use of phase cycling to remove any undesired contribution polluting our signal. These beams, labelled $A$, $B$, $C$ and $LO$, each carry a wavevector $\vec{k}_A$, $\vec{k}_B$, $\vec{k}_C$ and $\vec{k}_{LO}$.

Wavevector conservation imposes the desired signal to be generated at $\vec{k}_S = \vec{k}_B + \vec{k}_C - \vec{k}_A$ coinciding with the direction of the LO beam. In this geometry, beam A acts as the so-called conjugate beam since its wavevector contributes negatively to $k_S$ while all other beams are non-conjugate beams. When a two-dimensional spectrum is acquired with pulse sequences where the last beam to interact with the sample is non-conjugate, a one-quantum (1Q) spectrum is obtained. This family of measurements is so called since it probes transition from the ground state to states attainable by a single light-matter interaction and associated excited state absorption. The delay between the first and second pulse $t_{abs}$ is scanned and Fourier transformed to yield an absorptive frequency axis $\omega_{abs}$. The delay between the second and last pulse or population time $t_{pop}$ is not transformed since it generally only involves monotonically decaying population dynamics.

Rephasing or non-rephasing 1Q spectra are obtained when the first pulse interacting with the sample is respectively conjugated or non-conjugated. Summing the real part of both of these yields a total correlation or purely absortive 1Q spectra. This incarnation of a two-dimensional spectra yields the highest spectral resolution and clearly shows coherence and population transfer between spectral features as well as excited state absorption (ESA) as negative and positive features respectively. In figures S9 and S10, we present the real and imaginary parts of both rephasing and non-rephasing spectra corresponding to the total correlation spectra shown in this letter. Rephasing and non-rephasing spectra were not acquired simultaneously but during the same streak of experiments. An animated figure of the population time evolution of the total correlation spectra is also available online. The pump beam spectra was closely monitored during the acquisition of the data to ensure the experimental condition were exactly the same from acquisition to acquisition.

# S4 Dynamics

As previously mentionned, even though the population time generally does not involve oscillating coherences, it still carries information concerning the ultrafast population dynamics present within the system under study. In the main text, we invoke the presence of relaxation dynamics to infer the presence of cross-peaks buried within strong excited state absorption feature. We present in figure S11 the evolution during the population time of the ratio of two diagonal features in the total correlation spectra presented in the main text. As stated in the main text, excitations at low temperature quickly relax to lower energy states while they stabilize around some distribution at room temperature.

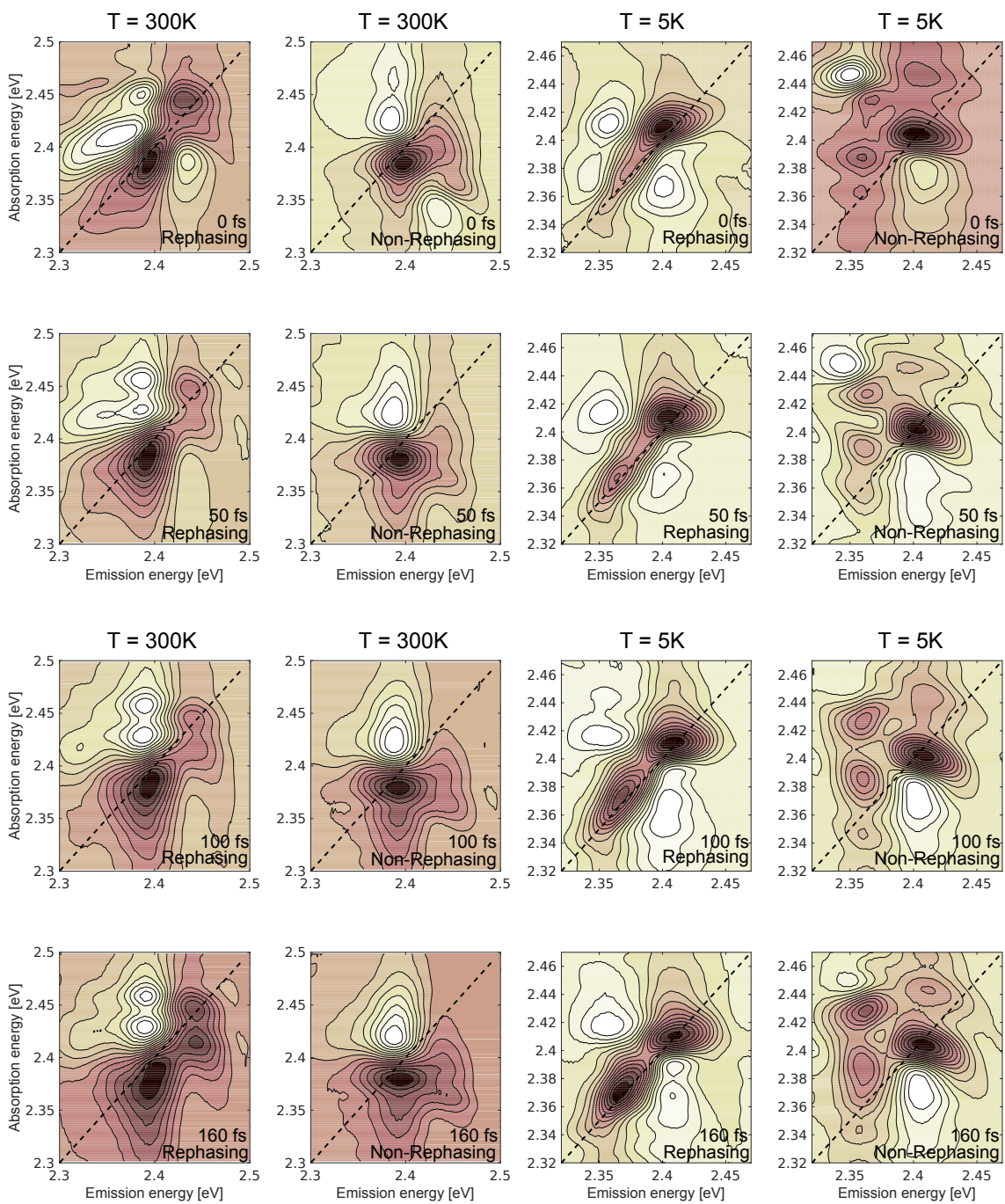

**Figure S9:** *Real part of rephasing and non-rephasing spectra for the corresponding 1Q spectra shown in this letter.*

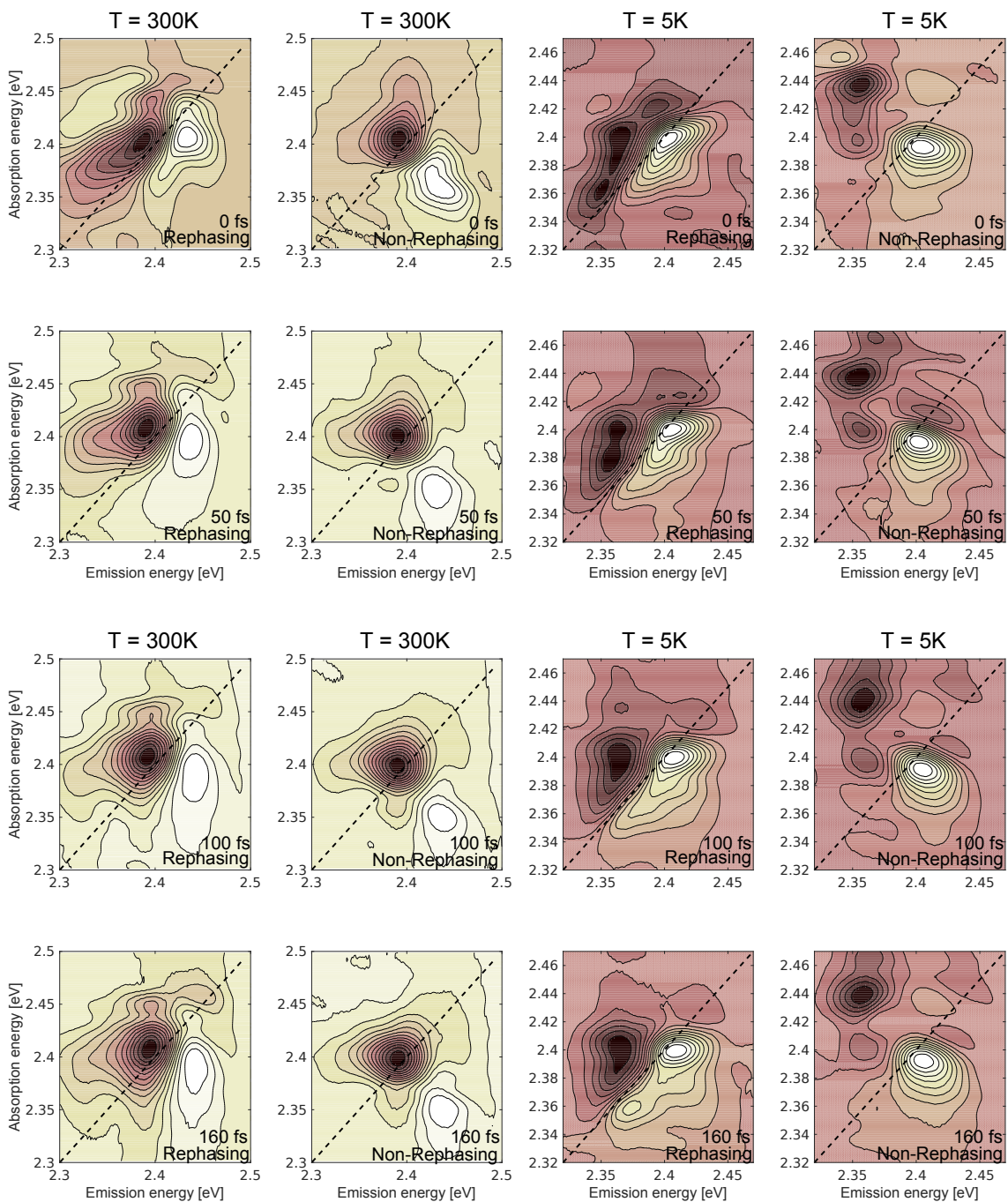

**Figure S10:** *Imaginary part of rephasing and non-rephasing spectra for the corresponding 1Q spectra shown in this letter.*

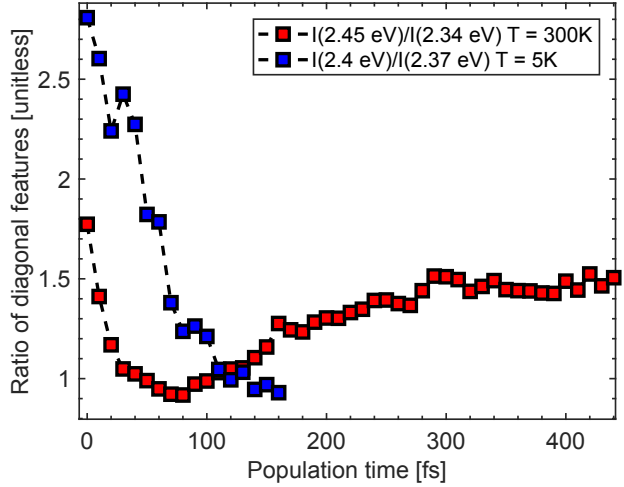

**Figure S11:** *Evolution with population time of the ratio of the diagonal features in total correlation spectra at low (blue squares) and room temperature (red squares). The chosen spectral position along the diagonal differs to accomodate redshift of the absorption spectra at low temperature.*

## S5   Pulse Characterisation and compression

Clean short pulses are essential to obtain a meaningful 2D spectrum devoid of any artefact arising from non time-ordered interactions. Before being sent into COLBERT, the input beam from the non-collinear optical parametric amplifer was precompressed using a folded single prism compressor [1]. For both characterisation and compression procedures, a 30 µm thin BBO crystal was placed at the sample position inside the cryostat before cooldown. All 4 beams are individually compressed using two iterations of chirp scan [2] followed by as many iterations of the MIIPS algorithm [3] as required for any detectable group delay dispersion (GDD) to disappear. Beams are routinely compressed under 50 fs$^2$ of GDD using this method.

One of the beams is then characterized (usually the local oscillator) in the same configuration using collinear frequency resolved optical gating (CFROG). The replica of the pulse under study is generated by applying a cosine modulation pattern on the phase pattern infered by the previous compression procedure [4]. Artefacts arising from the colinear nature of the measurement are then removed by a two-dimensional fourier transform [5]. The FROG trace obtained before performing the experiments presented in this letter is shown in figure S12b along with the extracted autocorrelation trace (a). The full width at half maximum (FWHM) of the latter indicates a pulse of 21 fs of temporal FWHM. To confirm that all other pulses are also correctly compressed, their cross-correlation with the beam characterized by CFROG is checked to be equal or under the previously obtained pulse width.

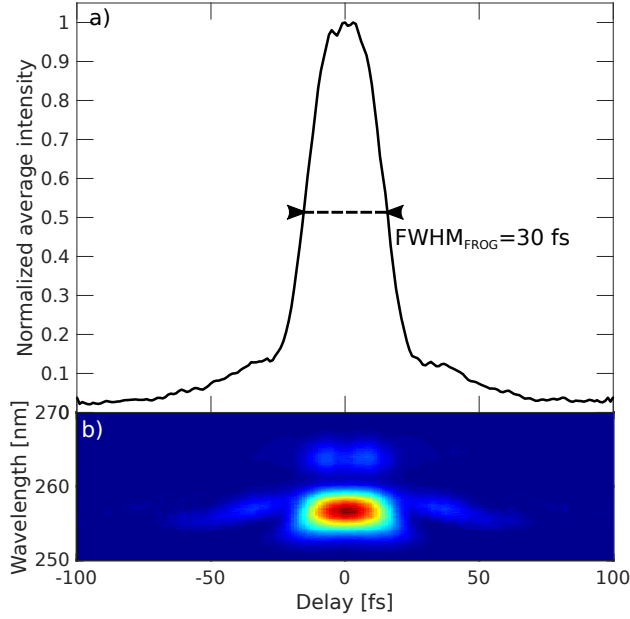

**Figure S12:** *(a) Spectral integration of the FROG trace presented in (b) for the local oscillator beam after compression*

# S6 COLBERT

To acquire the multidimensional spectra presented in the main text, we used a home-built COLBERT, a pulse-shapper based multidimensional spectrometer developped by the Nelson group. A detailled description of the inner workings of the spectrometer, its construction, calibration and operation can be found in an excellent review by the same group [6]. Since our implementation differs slightly , we present a sketch of it in figure S13. Differences include the use of an additionnal spectrometer to monitor the pump spectrum during the course of an experiment and the use of fiber optics to collect the weak signals involved in these experiments. The use of large core multimode fibers of short lengths allows the flexibilty of fiber optics to be used without the addition of non-linear artefacts in the acquired spectra. Additionnaly, since the signal and local oscillator acquire the same spectral phase during propagation in dispersive medium, the interference pattern (or relative phase) is unaffected.

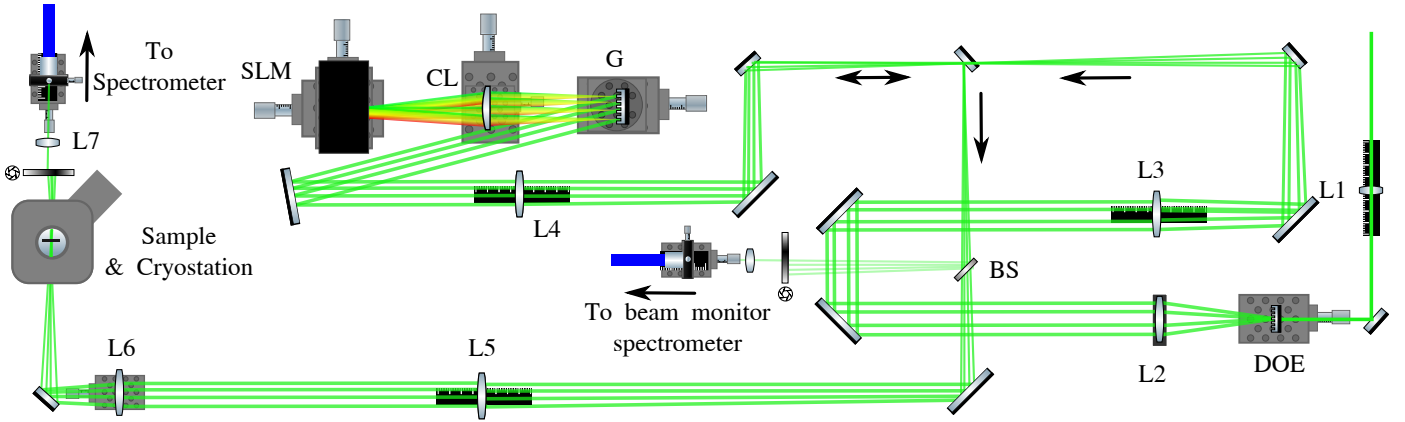

**Figure S13:** *A schematic of our home built multidimensional spectrometer. Lenses L1 to L7 relay the image the four beams generated by a diffractive optical element (DOE) to the sample. The diffractive pulse shaper composed of a spatial light modulator (SLM), a cylindrical lens (CL) and a grating (G) modulates in phase and amplitude the spectrum of each of the 4 beams. A beam sampler (BS) inserted after the pulse shapper collects a small fraction of the beams sent to the sample for continuous monitoring of the pump spectra.*



\end{document}